\documentclass{iopart}
\usepackage{epsfig}
\usepackage{subfigure}
\begin{document}
 
\article[]{}
  {Non-boost-invariant motion of dissipative and highly anisotropic fluid 
}  
 
\author{Radoslaw Ryblewski}
\address{The H. Niewodnicza\'nski Institute of Nuclear Physics,\\
  Polish Academy of Sciences,\\
  PL-31-342 Krak\'ow, Poland}
\ead{Radoslaw.Ryblewski@ifj.edu.pl}

\author{Wojciech Florkowski}
\address{Institute of Physics, Jan Kochanowski University, \\
         PL-25-406~Kielce, Poland, and \\
         The H. Niewodnicza\'nski Institute of Nuclear Physics,\\
         Polish Academy of Sciences,\\
         PL-31-342 Krak\'ow, Poland}
\ead{Wojciech.Florkowski@ifj.edu.pl}
   
\begin{abstract}
The recently formulated framework of anisotropic and dissipative hydrodynamics (ADHYDRO) is used to describe non-boost-invariant motion of the fluid created at the early stages of heavy-ion collisions. Very strong initial asymmetries of pressure are reduced by the entropy production processes. By the appropriate choice of the form of the entropy source we can describe isotropization times of about 1 fm, which agrees with the common expectations that already at such times the perfect-fluid hydrodynamics may be applied. Our previous results are generalized by including the realistic equation of state as the limit of the isotropization processes. 
\end{abstract}
\submitto{\JPG}
\maketitle

\section{Introduction}
\label{sect:intro}

In a recent paper we have introduced the new framework of highly-Anisotropic and strongly-Dissipative HYDROdynamics (ADHYDRO) \cite{Florkowski:2010cf}. This framework allows for studies of the systems which are locally highly anisotropic. Such situations happen at the early stages of relativistic heavy-ion collisions where, as suggested by many microscopic calculations \cite{Kovner:1995ja,Bjoraker:2000cf,El:2007vg}, the longitudinal pressure is much lower than the transverse pressure~\footnote{The directions are defined with respect to the beam axis}.  

In Ref. \cite{Florkowski:2010cf} our approach has been used to analyze the behavior of matter created in the central region of relativistic heavy-ion collisions, where the assumption of boost-invariance is acceptable. We have shown how our equations describe the isotropization of pressure and how the system approaches the regime governed by the perfect-fluid hydrodynamics. 

In this paper we relax the assumption of boost-invariance and consider a general one-dimensional expansion along the beam axis (at the early stages of collisions the longitudinal motion dominates). We derive equations describing non-boost-invariant motion of anisotropic fluid and show how the initial anisotropic behavior transforms smoothly into the locally isotropic expansion. 

In the second part of the paper we present a scheme implying that the isotropization transition leads to a desired equation of state in the perfect-fluid stage. In this way we construct an effective model which describes several stages of heavy-ion collisions in a very concise way: an early anisotropic phase (Glasma, color-flux tubes), transition to the isotropic phase with the QCD equation of state, perfect-fluid stage (sQGP), and the phase transition to the hadron gas.

We note that similar physical problems connected with the non-boost-invariant motion of the fluid formed at the early stages of relativistic heavy-ion collisions were studied in Refs. \cite{Satarov:2006jq,Bozek:2007qt}, see also \cite{Srivastava:1992gh,Srivastava:1992cg,Nagy:2007xn}. In Ref. \cite{Satarov:2006jq} the effects of the equation of state and the initial conditions were studied, while in Ref. \cite{Bozek:2007qt} the effects of the shear viscosity were taken additionally into account. In contrast to Refs. \cite{Satarov:2006jq,Bozek:2007qt} our approach concentrates on highly asymmetric stages of the evolution, see \cite{Bialas:2007gn,Bozek:2007di}. We also note that our formulation of ADHYDRO was followed by the paper by Martinez and Strickland \cite{Martinez:2010sc}, where a similar framework of anisotropic hydrodynamics was introduced. We discuss the connection between our work and Ref. \cite{Martinez:2010sc} in the Appendix.

\section{Anisotropic hydrodynamics}
\label{aniso-hyd}

In this Section we recapitulate the main assumptions of ADHYDRO, that have been recently introduced in \cite{Florkowski:2010cf}. Our starting point is the following form of the energy-momentum tensor,
\begin{eqnarray}
T^{\mu \nu} &=& \left( \varepsilon  + P_\perp\right) U^{\mu}U^{\nu} 
- P_\perp \, g^{\mu\nu} - (P_\perp - P_\parallel) V^{\mu}V^{\nu}. 
\label{Tmunudec}
\end{eqnarray}
Here $\varepsilon$, $P_\perp$, and $P_\parallel$ are the energy density, transverse pressure, and longitudinal pressure, respectively. For isotropic fluid, where the two pressures are equal, \mbox{$P_\perp=P_\parallel=P$}, we recover the form of the energy-momentum tensor of the perfect-fluid hydrodynamics. The four-vector $U^\mu$ in (\ref{Tmunudec}) represents the hydrodynamic flow, 
\begin{equation}
U^\mu = \gamma (1, v_x, v_y, v_z), \quad \gamma = (1-v^2)^{-1/2},
\label{Umu}
\end{equation}
while $V^\mu$ defines the direction of the longitudinal axis that plays a special role due to the initial geometry of the collision,
\begin{equation}
V^\mu = \gamma_z (v_z, 0, 0, 1), \quad \gamma_z = (1-v_z^2)^{-1/2}.
\label{Vmu}
\end{equation}
The four-vectors $U^\mu$ and $V^\mu$ satisfy the following normalization conditions:
\begin{eqnarray}
U^2 = 1, \quad V^2 = -1, \quad U \cdot V = 0.
\label{UVnorm}
\end{eqnarray}
In the local-rest-frame (LRF) of the fluid element we have $U^\mu = (1,0,0,0)$ and $V^\mu = (0,0,0,1)$, and the energy-momentum tensor has a simple diagonal structure~\footnote{The Lorentz transformation which leads from the center-of-mass frame, where Eqs. (\ref{Umu}) and (\ref{Vmu}) are valid, to LRF, where we have $U^\mu = (1,0,0,0)$ and $V^\mu = (0,0,0,1)$, consists of a Lorentz boost along the $z$-axis, a rotation around the $z$-axis, and a boost in the direction of the remaining transverse part of ${\bf v}$. In this way the initial longitudinal direction defined by the spatial part of $V^\mu$ is left unchanged.  },

\begin{equation}
T^{\mu \nu} =  \left(
\begin{array}{cccc}
\varepsilon & 0 & 0 & 0 \\
0 & P_\perp & 0 & 0 \\
0 & 0 & P_\perp & 0 \\
0 & 0 & 0 & P_\parallel
\end{array} \right).
\label{Tmunuarray}
\end{equation}
Hence, as expected, the formula (\ref{Tmunudec}) allows for different pressures in the longitudinal and transverse directions. Anisotropies of the form (\ref{Tmunuarray}) are present in the case where the particles are formed by the decays of color strings and in the theory of Color Glass Condensate \cite{Kovchegov:2009he,Kovchegov:2005ss,Krasnitz:2002mn} .

Besides the energy-momentum tensor (\ref{Tmunudec}), we introduce the entropy flux
\begin{eqnarray}
\sigma^{\mu} &=& \sigma U^{\mu},
\label{smudec}
\end{eqnarray}
where $\sigma$ is the entropy density. We assume that $\varepsilon$ and  $\sigma$ are functions of $P_\perp$ and $P_\parallel$. In particular, for massless partons the condition $T^\mu_{\,\,\,\mu}=0$ gives
\begin{equation}
\varepsilon = 2 P_\perp + P_\parallel.
\label{eos}
\end{equation}

The form of the energy-momentum tensor (\ref{Tmunudec}) resembles the form used in relativistic magnetohydrodynamics \cite{PhysRevE.47.4354,PhysRevE.51.4901}. In that case the anisotropy is induced by the presence of the magnetic field. At the early stages of heavy-ion collisions we have similar situation --- there exist strong color magnetic and electric longitudinal fields (Glasma following CGC \cite{Lappi:2006fp}) which polarize the medium. In this paper, however, we do not include explicitly the effects of the fields. Such analysis is left for a separate study. 

The space-time evolution of the system is governed by the equations expressing the energy-momentum conservation and the entropy growth,

\begin{eqnarray}
\partial_\mu T^{\mu \nu} &=& 0, \label{enmomcon} \\
\partial_\mu \sigma^{\mu} &=& \Sigma. \label{engrow}
\end{eqnarray}
The function $\Sigma$ represents the entropy source. The form of $\Sigma$ must be treated as an assumption that defines the dynamics of the anisotropic fluid. It is natural to assume that $\Sigma \geq 0$ and \mbox{$\Sigma = 0$} for \mbox{$P_\perp=P_\parallel$}. In this way, in the case where the two pressures are equal, the structure of the perfect-fluid hydrodynamics is recovered. 

We treat $\Sigma$ as a function of $P_\perp$ and $P_\parallel$. In this way, Eqs. (\ref{enmomcon}) and (\ref{engrow}) form a closed system of 5 equations for 5 unknown functions: three components of the fluid velocity, $P_\perp$, and $P_\parallel$. The projections of Eq. (\ref{enmomcon}) on $U_\nu$ and $V_\nu$ yield

\begin{eqnarray}
U^\mu \partial_\mu \varepsilon &=& - \left( \varepsilon+P_\perp \right) \partial_\mu U^\mu 
+ \left( P_\perp-P_\parallel \right) U_\nu  V^\mu \partial_\mu V^\nu, \label{enmomconU} \\
V^\mu \partial_\mu P_\parallel &=& - \left( P_\parallel-P_\perp \right) \partial_\mu V^\mu 
+ \left( \varepsilon+ P_\perp \right) V_\nu  U^\mu \partial_\mu U^\nu. \label{enmomconV}
\end{eqnarray}

\section{Anisotropy parameter $x$}
\label{sect:aniso-x}

Our earlier studies show \cite{Florkowski:2010cf,Florkowski:2008ag,Florkowski:2009sw} that it is useful to switch from $P_\perp$ and $P_\parallel$ to the two new variables: the entropy density  $\sigma$ and the anisotropy parameter $x$. In this case we may write

\begin{eqnarray}
\varepsilon &=&  \left(\frac{\pi^2 \sigma}{4 g_0} \right)^{4/3} R(x),
\label{epsilon2} 
\end{eqnarray}
\begin{eqnarray}
P_\perp &=&  \left(\frac{\pi^2 \sigma}{4 g_0} \right)^{4/3}
\left[\frac{R(x)}{3} + x R^\prime(x) \right],   
\label{PT2} 
\end{eqnarray}
\begin{eqnarray}
P_\parallel &=&  \left(\frac{\pi^2 \sigma}{4 g_0} \right)^{4/3} 
\left[\frac{R(x)}{3} - 2 x R^\prime(x) \right]. \label{PL2} \\ \nonumber
\end{eqnarray}
The function $R(x)$ is defined by the formula \cite{Florkowski:2009sw} \footnote{Note that for $x < 1$ the function $(\arctan\sqrt{x-1})/\sqrt{x-1}$ should be replaced by $(\hbox{arctanh}\sqrt{1-x})/\sqrt{1-x}$}

\begin{eqnarray}
R(x) = \frac{3\, g_0\, x^{-\frac{1}{3}}}{2 \pi^2} \left[ 1 + \frac{x \arctan\sqrt{x-1}}{\sqrt{x-1}}\right],
\label{Rx}  \\ \nonumber
\end{eqnarray}
and $g_0$ is the degeneracy factor connected with internal quantum numbers of particles that form the fluid. In our calculations we take $g_0=16$ --- the initial system is formed most likely from gluons.  The symbol $R'(x)$ denotes the derivative of $R(x)$ with respect to $x$, for $x=1$ we have $R'(1)=0$ and the two pressures are equal.

The physical interpretation of $x$ and details describing manipulations that lead to (\ref{epsilon2})--(\ref{Rx}) are given in \cite{Florkowski:2010cf,Florkowski:2009sw}. Here we only note that the ratio $P_\perp/P_\parallel$ is a monotonic  function of $x$ with $P_\perp > P_\parallel$ if $x > 1$ (to a good approximation $P_\perp/P_\parallel = x^{3/4}$, and $P_\perp = P_\parallel$ if $x = 1$).

\section{One-dimensional non-boost-invariant motion}
\label{sect:one-dim}

For a purely one dimensional motion we may introduce the following parameterizations:

\begin{equation}
U^{\mu} = \left(\cosh\vartheta(\tau,\eta),0,0,\sinh\vartheta(\tau,\eta)\right)
\label{Uv}
\end{equation}
and
\begin{equation}
V^{\mu} = \left(\sinh\vartheta(\tau,\eta),0,0,\cosh\vartheta(\tau,\eta)\right),
\label{Vv}
\end{equation}
where $\vartheta(\tau,\eta)$ is the fluid rapidity depending on the (longitudinal) proper time

\begin{equation}
\tau = \sqrt{t^2 - z^2}
\label{tau}
\end{equation} 
and the space-time rapidity 
\begin{eqnarray}
\eta = \frac{1}{2} \ln \frac{t+z}{t-z}. \label{eta} 
\end{eqnarray}
Equations (\ref{Uv}) and (\ref{Vv}) satisfy automatically the normalization conditions (\ref{UVnorm}). We also have
%
\begin{eqnarray}
U^\mu \partial_\mu &=& \cosh(\vartheta-\eta) \frac{\partial}{\partial \tau}
+ \frac{\sinh(\vartheta-\eta)}{\tau} \frac{\partial}{\partial \eta} .
\label{help1}
\end{eqnarray}
The expressions similar to (\ref{help1}) may be derived for $\partial_\mu U^\mu$, $V^\mu \partial_\mu$, and $\partial_\mu V^\mu$. Using them in Eqs. (\ref{enmomconU}) and (\ref{enmomconV}) gives

\begin{eqnarray}
\hspace{-2cm} \left[ \frac{\partial}{\partial \tau}
+ \frac{\tanh(\vartheta-\eta)}{\tau} \frac{\partial}{\partial \eta} \right] \varepsilon
&=&
-(\varepsilon + P_\parallel) 
\left[\tanh(\vartheta-\eta) \frac{\partial}{\partial \tau}
+  \frac{\partial}{\tau \partial \eta} \right] 
\vartheta,
\label{eq1} \\
\hspace{-2cm} \left[\tanh(\vartheta-\eta) \frac{\partial}{\partial \tau}
+  \frac{\partial}{\tau \partial \eta} \right] 
P_\parallel
&=&
-(\varepsilon + P_\parallel) 
\left[\frac{\partial}{\partial \tau}
+ \frac{\tanh(\vartheta-\eta)}{\tau} \frac{\partial}{\partial \eta} \right] 
\vartheta.
\label{eq2} 
\end{eqnarray}
Similarly, from Eq. (\ref{engrow}) one obtains 

\begin{equation}
\hspace{-2cm}  \left[ \frac{\partial}{\partial \tau}
+ \frac{\tanh(\vartheta-\eta)}{\tau} \frac{\partial}{\partial \eta} \right] \sigma
+ \sigma \left[\tanh(\vartheta-\eta) \frac{\partial}{\partial \tau}
+  \frac{\partial}{\tau \partial \eta} \right] \vartheta = \frac{\Sigma}{\cosh(\vartheta-\eta)}.
\label{eq3}
\end{equation}
The structure of  Eqs. (\ref{eq1})--(\ref{eq3}) is very much similar to the equations studied in Refs. \cite{Satarov:2006jq}. 

\section{Entropy source}

In the boost-invariant case the functions $\varepsilon$, $P_\parallel$, $P_\perp$, and $\sigma$ do not depend on $\eta$, and $\vartheta=\eta$. In this case Eq. (\ref{eq2}) is automatically fulfilled, while Eqs. (\ref{eq1}) and (\ref{eq3}) are reduced to 

\begin{eqnarray}
&& \frac{d \varepsilon}{d \tau} = -\frac{(\varepsilon + P_\parallel) }{\tau},
\label{eq1binv} \\
&& \frac{d \sigma}{d \tau} + \frac{\sigma}{\tau} = \Sigma.
\label{eq3binv}
\end{eqnarray}
Equations (\ref{eq1binv}) and (\ref{eq3binv}) were studied in \cite{Florkowski:2010cf} where the following form of the entropy source $\Sigma$ was used 
\begin{equation}
\Sigma = \frac{(1-\sqrt{x})^2}{\sqrt{x}} \frac{\sigma}{\tau_{\rm eq}}.
\label{ansatz1}
\end{equation} 
Here ${\tau_{\rm eq}}$ is a timescale parameter. The form (\ref{ansatz1}) guarantees that $\Sigma \geq 0$ and $\Sigma(\sigma,x=1)=0$. More arguments for such a particular form of $\Sigma$ are given in \cite{Florkowski:2010cf}. They connect $x$ with the ratio of the transverse and longitudinal temperatures. 

It may be shown that for small deviations from the equilibrium Eq. (\ref{ansatz1}) leads to quadratic dependence of the entropy source on the variable $\xi=1-x$ and this dependence is compatible with the Israel-Stewart theory (where the entropy production depends on the viscous stress squared) and with the Martinez-Strickland model \cite{Martinez:2010sc}. 

We have to stress, however, that the structure of the entropy source is an external input for the anisotropic hydrodynamics. It would be interesting to obtain any hints about $\Sigma$ for large $x$ from the microscopic models of particle production or from the AdS/CFT correspondence. 

\section{Isotropization}

If $\varepsilon$ and $P_\parallel$ are expressed in terms of $\sigma$ and $x$, Eqs. (\ref{eq1})--(\ref{eq3}) become three equations for three unknown functions: $\sigma$, $x$, and $\vartheta$:

\begin{eqnarray}
\hspace{-2cm} \frac{4}{3 \sigma} \left[ \frac{\partial}{\partial \tau}
+ \frac{\tanh(\vartheta-\eta)}{\tau} \frac{\partial}{\partial \eta} \right] \sigma
&=&
- \frac{R'(x)}{R(x)}
\left[ \frac{\partial}{\partial \tau}
+ \frac{\tanh(\vartheta-\eta)}{\tau} \frac{\partial}{\partial \eta} \right] x \nonumber \\
& & - \left( 1+h(x) \right) 
\left[\tanh(\vartheta-\eta) \frac{\partial}{\partial \tau}
+  \frac{\partial}{\tau \partial \eta} \right] \vartheta,
\label{eq1sigx} 
\end{eqnarray}

\begin{eqnarray}
\hspace{-2cm} \frac{4}{3 \sigma} 
\left[\tanh(\vartheta-\eta) \frac{\partial}{\partial \tau}
+  \frac{\partial}{\tau \partial \eta} \right]
 \sigma
&=&
- \left( \frac{R'(x)}{R(x)}+\frac{h'(x)}{h(x)} \right)
\left[\tanh(\vartheta-\eta) \frac{\partial}{\partial \tau}
+  \frac{\partial}{\tau \partial \eta} \right] x \nonumber \\
& & - \frac{1+h(x)}{h(x)}  \left[ \frac{\partial}{\partial \tau}
+ \frac{\tanh(\vartheta-\eta)}{\tau} \frac{\partial}{\partial \eta} \right]
 \vartheta,
\label{eq2sigx} 
\end{eqnarray}

\begin{equation}
\hspace{-2cm}  \left[ \frac{\partial}{\partial \tau}
+ \frac{\tanh(\vartheta-\eta)}{\tau} \frac{\partial}{\partial \eta} \right] \sigma
+ \sigma \left[\tanh(\vartheta-\eta) \frac{\partial}{\partial \tau}
+  \frac{\partial}{\tau \partial \eta} \right] \vartheta = \frac{\Sigma}{\cosh(\vartheta-\eta)}.
\label{eq3sigx}
\end{equation}
Here we have introduced the ratio of the longitudinal pressure and energy density, which equals 1/3 for the isotropic system with $x=1$,

\begin{equation}
h(x) = \frac{P_\parallel}{\varepsilon}, \quad h(1) = \frac{1}{3}.
\end{equation}
Substituting (\ref{eq3sigx}) into (\ref{eq1sigx}) and using the relation 

\begin{equation}
1 + h(x) = \frac{4}{3} - 2 x \frac{R'(x)}{R(x)},
\label{oneplush}
\end{equation}
we find

\begin{eqnarray}
\hspace{-2cm} \left[ \frac{\partial}{\partial \tau}
+ \frac{\tanh(\vartheta-\eta)}{\tau} \frac{\partial}{\partial \eta} \right] x
&=&
2 x 
\left[\tanh(\vartheta-\eta) \frac{\partial}{\partial \tau}
+  \frac{\partial}{\tau \partial \eta} \right] \vartheta 
-\frac{4 H(x) }{3 \tau_{\rm eq} \cosh(\vartheta-\eta)}.
\label{eqHsigx}  \\ \nonumber
\end{eqnarray}
Here we use the notation introduced in \cite{Florkowski:2010cf},

\begin{equation}
H(x) = \frac{R(x)}{R'(x)} \frac{(1-\sqrt{x})^2}{\sqrt{x}}.
\label{Hofx}
\end{equation}
In the case where $\vartheta(\tau,\eta) \approx \eta$, Eq. (\ref{eqHsigx}) may be approximated by the ordinary differential equation derived in \cite{Florkowski:2010cf}

\begin{equation}
\frac{dx}{d\tau} = \frac{2x}{\tau} -\frac{4H(x)}{3 \tau_{\rm eq}}.
\label{xoftau}
\end{equation}
Thus, in this case we may immediately use our previous results to argue that $x \to 1$ for $\tau \gg \tau_{\rm eq}$. If the condition $\vartheta(\tau,\eta) \approx \eta$ is not fulfilled, one should study Eqs. (\ref{eq1sigx})--(\ref{eq3sigx}) numerically.

\section{Initial conditions}

The initial conditions for a non-boost-invariant evolution are defined by three functions: $\sigma(\tau_0,\eta)$, $\vartheta(\tau_0,\eta)$, and $x(\tau_0,\eta)$, where $\tau_0$ is the initial proper time. In the numerical calculations we assume \mbox{$\tau_0 =$ 0.2 fm}. The initial entropy profile is taken in the form \cite{Hirano:2002ds,Bozek:2009ty}
\begin{eqnarray}
\sigma(\tau_0, \eta) = \sigma_0 \exp\left[
-\frac{(|\eta|-\Delta\eta)^2}{2 (\delta\eta)^2} \; \theta(|\eta|-\Delta\eta)
\right],
\label{init_sigma}
\end{eqnarray}
where $\theta$ is the step function, the parameter $\Delta\eta$ defines the half width of the initial  plateau in spacetime rapidity, and $\delta\eta$ defines the half width of the Gaussian tails on both sides of the plateau. By the appropriate changes of the parameters $\Delta\eta$ and $\delta\eta$ we may vary between the boost-invariant-like and Gaussian-like initial conditions. To match the rapidity distribution measured by BRAHMS \cite{Bearden:2004yx} we use the values: $\Delta\eta=1$ and $\delta\eta=1.3$. The value of the initial central entropy density is obtained from the condition that the initial energy density is 100 GeV/fm$^3$,
\begin{equation}
\varepsilon_0 = 100 \, \hbox{GeV/fm$^3$} = \left(\frac{\pi^2 \sigma_0}{4 g_0} \right)^{4/3} R(x_0).
\label{eps0}
\end{equation}
Here, $x_0$ is the initial value of the anisotropy parameter at $\eta=0$. In the present calculations we assume that the initial profile of $x$ is constant, 
\begin{equation}
x(\tau_0, \eta) = x_0 = 100.
\label{init_vartheta}
\end{equation}
This value of $x_0$ is somewhat arbitrary but it expresses the opinion that the initial longitudinal pressure is much smaller than the initial transverse pressure. 

The value of the timescale parameter $\tau_{\rm eq}$ has been set equal to 0.25 fm. Finally, in agreement with other  hydrodynamic calculations which address the problem of the longitudinal expansion, we take the initial fluid rapidity profile in the form
\begin{equation}
\vartheta(\tau_0, \eta) = \eta.
\label{init_vartheta}
\end{equation}
Of course, for the boost-invariant motion we have always $\vartheta=\eta$. In our case, due to the presence of the longitudinal gradients determined by the entropy profile (\ref{init_sigma}), we expect $|\vartheta| \geq |\eta|$.

\begin{figure}[t]
\begin{center}
\subfigure{\includegraphics[angle=0, width=0.7\textwidth]{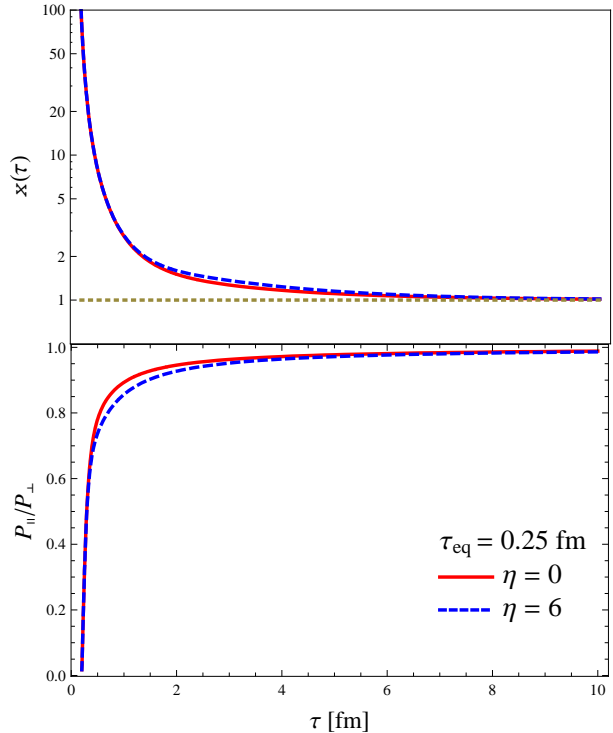}} 
\end{center}
\caption{Upper part: Time evolution of the anisotropy parameter $x$ for two different values of the space-time rapidity: $\eta=0$ (solid line) and $\eta=6$ (dashed line). Lower part: time evolution of the ratio $P_\parallel/P_\perp$ for $\eta=0$ (solid line) and $\eta=6$ (dashed line).  }
\label{fig:XS}
\end{figure}

\begin{figure}[t]
\begin{center}
\subfigure{\includegraphics[angle=0,width=0.7\textwidth]{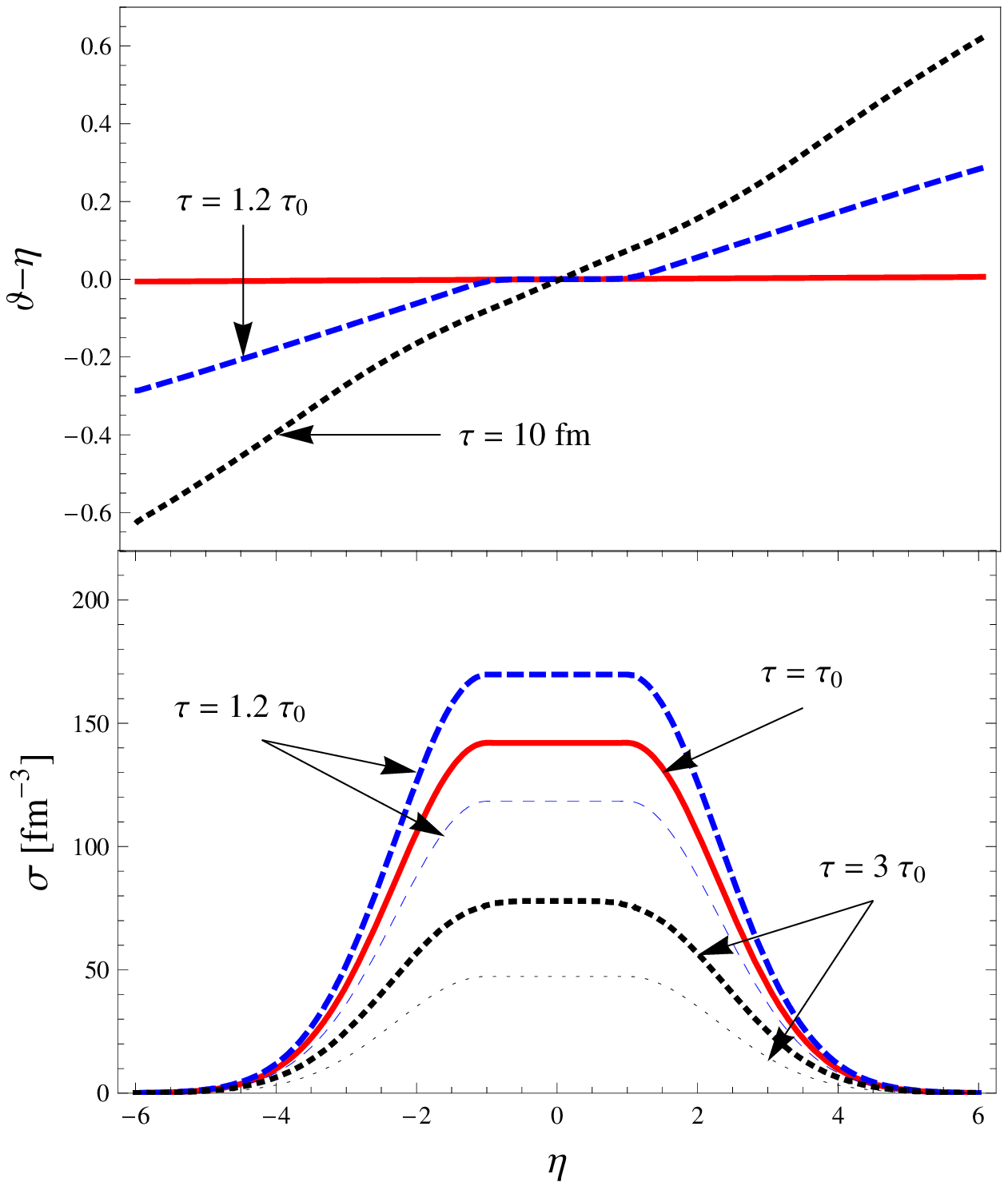}} 
\end{center}
\caption{ Upper part: difference between the fluid rapidity $\vartheta$ and the space-time rapidity $\eta$ for $\tau=\tau_0$ (solid line), $\tau = 1.2 \tau_0$ (dashed line), and \mbox{$\tau = 10$ fm} (dotted line). Lower part:  profiles of the entropy density plotted for different values of the proper time. The results of our numerical calculations are represented by the thick lines. The results obtained from the simple scaling $\sigma(\tau,\eta)=\sigma(\tau_0,\eta) \tau_0/\tau$ are represented by the thin lines. }
\label{fig:TP}
\end{figure}

\section{Results}

The results of our numerical calculations with the initial conditions specified in the previous Section are presented in Figs. \ref{fig:XS} and \ref{fig:TP}. The upper part of Fig. \ref{fig:XS} shows the time evolution of the anisotropy parameter $x$ for two different values of the space-time rapidity: $\eta=0$ (solid line) and $\eta=6$ (dashed line). We observe the initial rapid decrease of $x$ (from 100 to about 5) followed by a longer evolution where $x$ approaches unity. This behavior is very weakly dependent on rapidity and very much similar to that observed previously in the boost-invariant case. 

The lower part of Fig. \ref{fig:XS} shows the time evolution of the ratio $P_\parallel/P_\perp$. The time dependence of this ratio reflects the time dependence of the anisotropy parameter $x$. Again, we observe a rapid change at the very beginning of the evolution that is followed by a more moderate approach to unity. For \mbox{$\tau > 2$ fm} the system is practically isotropic. Clearly, by changing the value of $\tau_{\rm eq}$ we may speed up or slow down the process of isotropization. The most common perfect-fluid description of the data requires that the stage described by 
perfect hydrodynamics starts at about 1 fm or even earlier. Therefore, we think that the time evolution obtained in our approach may be regarded as realistic if the systems becomes isotropic at about 1 fm. The results shown in Figs. \ref{fig:XS} and \ref{fig:TP} suggest that this condition may be achieved if $\tau_{\rm eq} \leq 0.1$ fm. Such extremely short time scales indicate very strong interactions in the early anisotropic plasma. 

The upper part of Fig. \ref{fig:TP} shows the difference between the fluid rapidity $\vartheta$ and the space-time rapidity $\eta$ for $\tau=\tau_0$ (solid line), $\tau = 1.2 \tau_0$ (dashed line), and \mbox{$\tau = 10$ fm} (dotted line). In the boost-invariant case $\vartheta=\eta$. In the present calculation, due to the extra pressure gradients determined by the entropy profile, the values of the fluid rapidity become larger than the space-time rapidity (for $\eta > 0$). 

The lower part of Fig. \ref{fig:TP} shows the entropy density profiles in rapidity for different values of the proper time. The solid line shows the initial entropy profile (\ref{init_sigma}) with the central entropy density $\sigma_0$ determined by the condition (\ref{eps0}). The thick dashed line shows the entropy density profile for the proper time $\tau=1.2\tau_0$. Since the initial dynamics is dissipative, the entropy is produced in the early stage, and the thick dashed line is placed above the solid line. For comparison, the thin dashed line shows the entropy profile expected from the boost-invariant isentropic expansion where $\sigma(\tau,\eta)=\sigma(\tau_0,\eta) \tau_0/\tau$. The thin dashed line is placed below the solid line, since the entropy density decreases in this case due to the longitudinal expansion of the fluid. By comparing the maximal values of the thick and dashed lines we can make the estimate of the relative entropy production due to the dissipative effects present in our approach. From this simple calculation we find that the entropy increases by about 70\%.

\begin{figure}[t]
\begin{center}
\subfigure{\includegraphics[angle=0,width=0.7\textwidth]{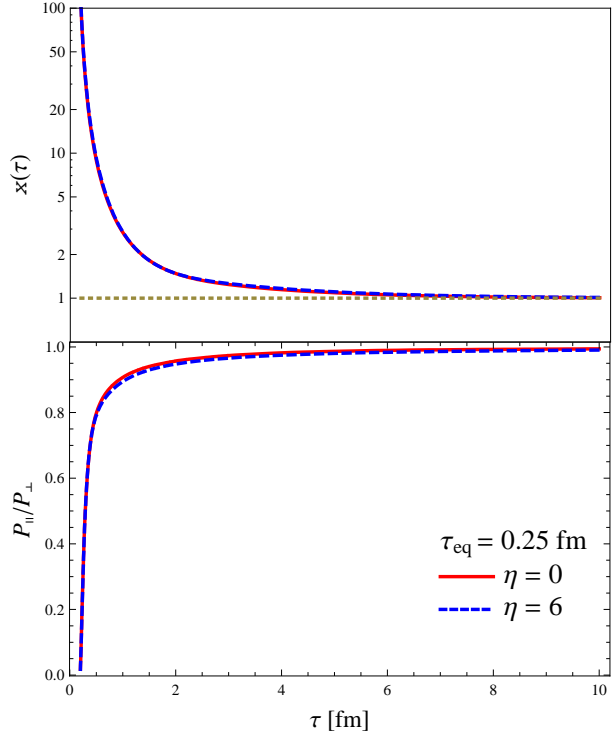}}
\end{center}
\caption{The same as Fig. \ref{fig:XS} but with Eqs. (\ref{epsilon2})--(\ref{PL2}) replaced by Eqs. (\ref{epsilon2qgp})--(\ref{PL2qgp}). }
\label{fig:XSeos}
\end{figure}

\section{Transition to quark-gluon plasma}

The concept of ADHYDRO is motivated by the microscopic picture of the anisotropic parton distribution function  \cite{Florkowski:2010cf,Florkowski:2009sw}. As the system becomes isotropic, the parton distribution becomes also isotropic (locally, in the momentum space). This, however, does not automatically mean that the isotropic system is described by the realistic equation of state. The Boltzmann distribution used in Refs. \cite{Florkowski:2010cf,Florkowski:2009sw} may be inadequate for description of the strongly interacting quark-gluon plasma. To connect the process of isotropization with the process of formation of the equilibrated quark-gluon plasma we may consider the following ansatz for the energy density and pressure

\begin{eqnarray}
\varepsilon(\sigma, x) &=&  \frac{\varepsilon_{\rm qgp}(\sigma)}{3} \frac{\pi^2}{g_0} R(x),  
\label{epsilon2qgp} \\
P_\perp(\sigma, x) &=&  P_{\rm qgp}(\sigma) \frac{\pi^2}{g_0} \left[ \frac{R(x)}{3} + x R'(x) \right].   
\label{PT2qgp} \\
P_\parallel(\sigma, x) &=&  P_{\rm qgp}(\sigma) \frac{\pi^2}{g_0} \left[ \frac{R(x)}{3} - 2 x R'(x) \right].  
\label{PL2qgp}
\end{eqnarray}
Here, the functions $\varepsilon_{\rm qgp}(\sigma)/3$ and $P_{\rm qgp}(\sigma)$ describe the realistic equation of state constructed in \cite{Chojnacki:2007jc}. They appear in Eqs. (\ref{epsilon2qgp})--(\ref{PL2qgp}) in the very much analogous way as the term $\sigma^{4/3}$  in Eqs. (\ref{epsilon2})--(\ref{PL2}). Although  $\varepsilon_{\rm qgp}(\sigma)$ and $P_{\rm qgp}(\sigma)$ were calculated for equilibrium, they are used by us to calculate the energy density and pressure for an arbitrary state characterized by the entropy $\sigma$ and the anisotropy parameter $x$.  In order to take into account the effects connected with the anisotropy, we multiply $\varepsilon_{\rm qgp}(\sigma)/3$ and $P_{\rm qgp}(\sigma)$ by the same combinations of the function $R(x)$ as in Eqs. (\ref{epsilon2})--(\ref{PL2}). The factor $\pi^2/g_0$ has been introduced to guarantee that  $\varepsilon(\sigma, x=1) =  \varepsilon_{\rm qgp}(\sigma)$, and similarly $P_\parallel(\sigma, x=1)=P_\perp(\sigma, x=1) =  P_{\rm qgp}(\sigma)$.

\begin{figure}[t]
\begin{center}
\subfigure{\includegraphics[angle=0,width=0.7\textwidth]{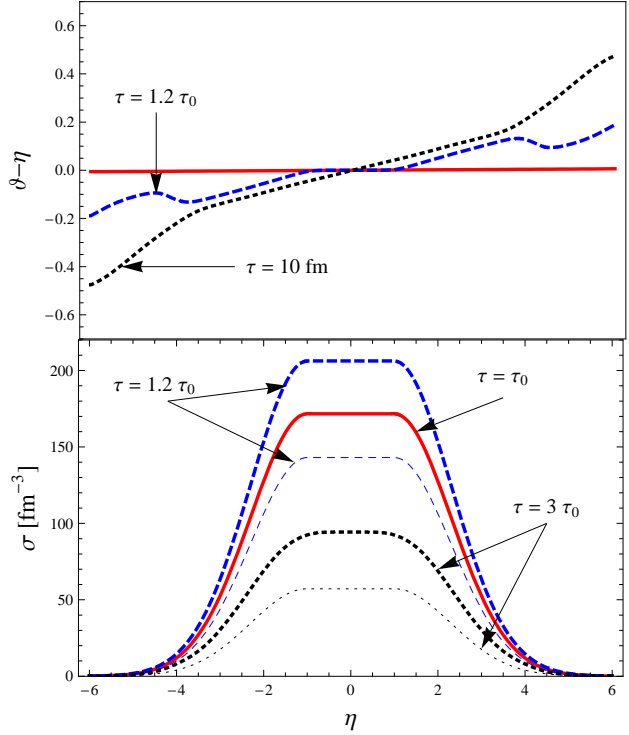}}
\end{center}
\caption{The same as Fig. \ref{fig:TP} but again with Eqs. (\ref{epsilon2})--(\ref{PL2}) replaced by Eqs. (\ref{epsilon2qgp})--(\ref{PL2qgp}).   }
\label{fig:TPeos}
\end{figure}

We should admit that Eqs. (\ref{epsilon2qgp})--(\ref{PL2qgp}) have no direct microscopic explanation. The main argument for using this form is that it has two attractive limits and may be used as an interpolating formula. At the very early stage, the system consists most likely of massless partons and Eqs. \mbox{(\ref{epsilon2qgp})--(\ref{PL2qgp})} approach Eqs. \mbox{(\ref{epsilon2})--(\ref{PL2})}, since the realistic equation of state approaches the Stefan-Boltzmann limit (although deviations from this limit are noticeable). On the other hand, if the system becomes isotropic, $x \to 1$, $R^\prime(x) \to 0$, and Eqs. \mbox{(\ref{epsilon2qgp})--(\ref{PL2qgp})}  are reduced to the equation of state used in standard hydrodynamics. From the statistical point of view, the initial system of partons is always treated classically and $R(x)$ is given by (\ref{Rx}). Inclusion of quantum statistics and consequences of other model assumptions will be discussed in a separate paper.

In Figs. \ref{fig:XSeos} and \ref{fig:TPeos} we show the results of the numerical calculations where Eqs. (\ref{epsilon2qgp})--(\ref{PL2qgp}) were substituted in Eqs. (\ref{eq1})--(\ref{eq3}). By this substitution we have obtained the system of three equations for $\sigma$, $x$, and $\vartheta$, which is quite analogous to Eqs. (\ref{eq1sigx})--(\ref{eq3sigx}). The values of the initial parameters and the initial profiles have not been changed except for the initial value of the entropy density, which has been modified to keep the initial energy density unchanged,
\begin{equation}
\varepsilon_0 = 100 \, \hbox{GeV/fm$^3$} = \varepsilon_{\rm qgp}(\sigma_0) \frac{\pi^2}{g_0} \frac{R(x_0)}{3}.
\label{eps0qgp }
\end{equation}

The overall space-time evolution of the system shown in Figs. \ref{fig:XSeos} and \ref{fig:TPeos} resembles very much the evolution shown earlier in Figs. \ref{fig:XS} and \ref{fig:TP}. In particular, the time changes of $x$ and the ratio of pressures are practically the same. The small differences are visible in the evolution of the entropy density (mainly due to a different normalization caused by fixing the initial energy density rather than the entropy density) and the fluid rapidity. The difference $\vartheta-\eta$ becomes now a more complicated function of $\eta$, the effect noticed also in Ref. \cite{Bozek:2007qt} and connected with the use of the equation of state from Ref. \cite{Chojnacki:2007jc}.

Clearly, an attractive feature of using Eqs. (\ref{epsilon2qgp})--(\ref{PL2qgp}) instead of Eqs. (\ref{epsilon2})--(\ref{PL2}) is that the system reaching equilibrium is described by the realistic QCD equation of state. In this way, the intermediate hydrodynamic evolution agrees with our expectations and no extra construction (like, e.g., the Landau matching condition used in \cite{Ryblewski:2010tn}) is necessary to change from one thermodynamic regime to another one.  On the other hand, the connection of Eqs. (\ref{epsilon2qgp})--(\ref{PL2qgp}) to any microscopic picture is lacking, since the realistic QCD equation of state cannot be simply interpreted in a quasi-particle picture. Another attractive feature Eqs. (\ref{epsilon2qgp})--(\ref{PL2qgp}) is that they include the phase transition from the quark-gluon plasma to the hadron gas. Thus, using  Eqs. (\ref{epsilon2qgp})--(\ref{PL2qgp}) we may describe many stages of heavy-ion collisions in a single approach.

\section{Conclusions}

Our results confirm very attractive features of ADHYDRO as an effective model for early stages of heavy-ion collisions. In this paper we have shown that the tendency  to approach isotropy is not changed by non-boost-invariant profiles. Very strong initial asymmetries of pressure are reduced by the entropy production processes. By the appropriate choice of the form of the entropy source and the value of the timescale parameter, we have succeeded to reproduce the realistic times for the isotropization expected in heavy-ion collisions. We have also generalized our previous result by including the realistic equation of state as the limit of the isotropization processes. 

From the practical point of view, our formalism based on the anisotropic distribution function allows for the determination of the space-time evolution of color-neutral anisotropic distributions, which may be used, 
for example, as background distributions in the analysis of the plasma instabilities.

The results obtained within ADHYDRO in this paper, and also in our previous publication,  open perspectives for further applications of the model. In particular, the model may be applied in the 2+1  case (longitudinal boost-invariance without cylindrical symmetry) or in the general 3+1 case. In the 2+1 case it may be treated as a generalization of the framework where transverse hydrodynamics is combined with the perfect-fluid hydrodynamics by the Landau matching conditions \cite{Ryblewski:2010tn}. In ADHYDRO the change from the system where only transverse degrees of freedom are thermalized to the full local equilibrium is described in the continuous way.

\section{Appendix}

Our formulation of ADHYDRO was followed by the paper by Martinez and Strickland \cite{Martinez:2010sc} where similar ideas appeared. It is interesting to show that the two approaches agree for small deviations from equilibrium in the boost-invariant case. In the following, it is useful to use the variable
\begin{equation}
\xi = x-1.
\label{xi}
\end{equation}
Equation (20) from \cite{Martinez:2010sc} connects $\xi$ and $p_{\rm hard}$,
\begin{equation}
\frac{1}{1+\xi} \frac{d\xi}{d\tau} - \frac{2}{\tau} - \frac{6}{p_{\rm hard}} \frac{dp_{\rm hard}}{d\tau}
= 2 \Gamma \left[ 1- {\cal R}^{3/4}(\xi) \sqrt{1+\xi} \right].
\label{MS20}
\end{equation}
Here, $p_{\rm hard}$ is related to the average momentum in the parton distribution function and $\cal R(\xi)$ is defined by the formula
\begin{eqnarray}
{\cal R}(\xi) = \frac{1}{2} \left[ \frac{1}{1+\xi} + \frac{\arctan\sqrt{\xi}}{\sqrt{\xi}}\right].
\label{calR}  \\ \nonumber
\end{eqnarray}
For $|\xi| < 1$ we expand ${\cal R}(\xi)$ around zero. Keeping the leading terms in $\xi$ one obtains from (\ref{MS20})
\begin{equation}
\frac{1}{1+\xi} \frac{d\xi}{d\tau} - \frac{2}{\tau} - \frac{6}{p_{\rm hard}} \frac{dp_{\rm hard}}{d\tau}
= -\frac{ \Gamma \xi^2}{15}.
\label{MS20limit}
\end{equation}

In order to connect this result with our approach, we relate the non-equilibrium entropy density $\sigma$ to $p_{\rm hard}$ \cite{Martinez:2010sc},
\begin{equation}
\sigma = A p_{\rm hard}^3 x^{-1/2}.
\label{sig-non-iso}
\end{equation}
Here $A$ is an irrelevant constant. Inserting Eq. (\ref{sig-non-iso}) into Eq. (\ref{eq3binv}) and using our ansatz for $\Sigma$ we find
\begin{equation}
-\frac{1}{x} \frac{dx}{d\tau} + \frac{2}{\tau} + \frac{6}{p_{\rm hard}} \frac{dp_{\rm hard}}{d\tau}
= \frac{2}{\tau_{\rm eq}} \frac{(1-\sqrt{x})^2}{\sqrt{x}} \approx \frac{(x-1)^2}{2 \tau_{\rm eq} }.
\label{MS20comp}
\end{equation}
Thus the two approaches are consistent if we set
\begin{equation}
\Gamma = \frac{15}{2 \tau_{\rm eq} } .
\label{Gammataueq}
\end{equation}

The consistency condition (\ref{Gammataueq}) may be further used in Eq. (24a) of Ref. \cite{Martinez:2010sc} to show that in the limit $|\xi| < 1$ it agrees with our fundamental expression (\ref{xoftau}). To do so, we rewrite Eq. (24a) in the analogous form, namely 
\begin{equation}
\frac{dx}{d\tau} = \frac{2x}{\tau} -\frac{4 \Gamma H_{MS}(x)}{3}.
\label{xoftauMS}
\end{equation}
\begin{figure}[t]
\begin{center}
\includegraphics[angle=0,width=0.65\textwidth]{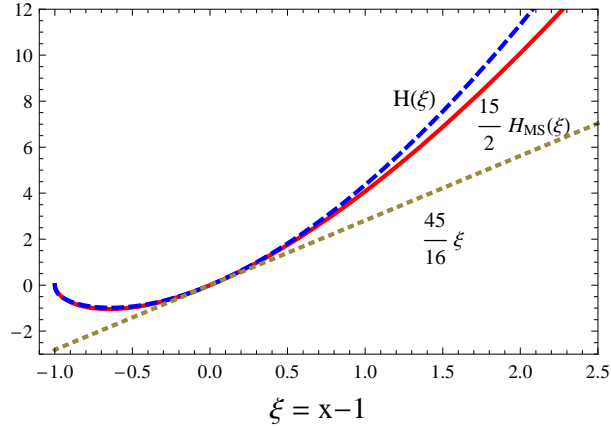} 
\end{center}
\caption{Functions $H(\xi)$ and $(15/2) H_{MS}(\xi)$ used in the present paper and in Ref. \cite{Martinez:2010sc}. For small $\xi$ they are both well approximated by the linear form $(45/16) \xi$. }
\label{fig:M2}
\end{figure}
In Fig. \ref{fig:M2} we show our function $H(\xi)$ and the function $H_{MS}(\xi)$ ($\xi=x-1$) multiplied by the factor $15/2$. The two functions agree for small values of $\xi$, hence again the consistency is achieved between the two approaches if the condition (\ref{Gammataueq}) is applied. Since the authors of Ref.  \cite{Martinez:2010sc} showed that their approach is reduced to the 2nd order viscous hydrodynamic equations of Israel and Stewart in the leading order in $\xi$, the comparison showed in this Appendix shows that our approach is also reduced to viscous hydrodynamics in the considered case. 

We note that $H(\xi)$ and  $(15/2) H_{MS}(\xi)$ differ for large values of $\xi$. In this region there are no direct hints about the behavior of matter from the underlying kinetic theory and the form of the function $H(\xi)$ is not constrained. It may be postulated and eventually verified by making comparisons of the model results with the experimental data. Such analyses represent the most interesting applications of ADHYDRO.

\section{Acknowledgments} 

This work was supported in part by the MNiSW grants No. N N202 288638 and N N202 263438.

\bigskip


\end{document}